# *SU(3)* analysis of nonfactorizable contributions to Bottom mesons decays

Maninder Kaur[*] and R. C. Verma[†]
*Department of Physics, Punjabi University,*
*Patiala – 147002, India.*

Abstract

This paper is the extension of our previous work entitled "Searching a systematics for nonfactorizable contributions to $B^-$ and $\bar{B}^0$ hadronic decays". In order to realize the full impact of isospin analysis, and to relate decays of strange bottom meson with those of nonstrange bottom mesons, we generalize it to the *SU(3)* flavor symmetry to investigate the nonfactorizable contributions to CKM-enhanced and CKM-suppressed decays. We start with expressing total weak decay amplitude as sum of the factorizable and nonfactorizable parts, then obtain the factorizable part of the decay amplitude numerically using the known meson decay constants and relevant form-factors, then express the nonfactorizable parts in terms of *SU(3)*-reduced matrix elements. Using measured branching fractions of a few CKM-favored modes, we fix the reduced matrix elements, and predict branching ratios of the remaining CKM-favored decays and all CKM-suppressed $\bar{B} \to PP$ decays involving $b+u \to c+d/s$ processes. We find that the measured branching fractions agree well with our results, and other predicted values may be tested in the future experiments.

* maninderkaur@pbi.ac.in
† rcverma@gmail.com

# I. INTRODUCTION

Experimentally, extensive measurements have been made on the weak decays of charm and bottom flavor hadrons [1]. The study of *B*-meson decays plays a crucial role in the testing of the Standard Model (SM) and measurements of its parameters. The SM is reasonably successful in explaining the leptonic and semileptonic decays, but nonleptonic decays are notoriously the most challenging due to the strong interaction interference with the weak interactions responsible for these decays [2-8]. The confinement itself is the manifestation of strong force, even though no stable mesons exist. The weak hadronic decays are often seen as a simplest testing ground of QCD. In fact, understanding of the decays gets more complicated as the produced hadrons in the weak hadronic decays can participate in the Final State Interactions (FSI) [9-10] caused by the strong interactions at hadronic level. Therefore, analysis of weak hadronic decays require phenomenological treatment for which symmetry principles and quark models are often employed for exploring the dynamics involved.

Even the weak interaction vertex itself is also affected through gluons-exchange among the quarks involved. At *W*-mass scale, hard gluons exchange effects are calculable using the perturbative QCD. Usually, factorization of weak matrix elements is performed in terms of certain form-factors and decay constants. Besides these high energy gluon exchanges, there exist possibility of soft gluon exchanges around the *W*-vertex, which generate the nonfactorizable contributions in the weak matrix elements [11-12]. The nonfactorizable terms may appear for several reasons, like soft gluon exchange, FSI rescattering effects [13-14]. The rescattering effects on the outgoing mesons have been studied in detail for bottom meson decays [15-16]. Besides that, flavor *SU(3)* symmetry and Factorization Assisted Topological (FAT) approach have been employed for the study of such nonfactorizable contributions, as they have the advantage of absorbing various kinds of contributions lump- sum in terms of a few parameters, to be fixed empirically [17-18]. Extensive work has also been conducted to treat nonfactorizable contributions such as QCD factorization approach based on collinear factorization theorem [19] and the perturbative QCD factorization approach [20-21]. Unfortunately, it is not straightforward to calculate such terms, which are nonperturbative in nature, and require empirical data to investigate their behavior.

In the naïve factorization scheme, the nonfactorizable contribution for the decay amplitudes was totally ignored and the two QCD coefficients $a_1$ and $a_2$ are fixed from the experimental data [22-24]. Initially, data on branching fractions of $D \to \bar{K}\pi$ decays seemed to require, $a_1 \approx c_1 = 1.26, \quad a_2 \approx c_2 = -0.51,$ leading to destructive interference between color-favored (CF) and color-suppressed (CS) processes for $D^+ \to \bar{K}^0\pi^+$, thereby implying the $N_c \to \infty$ limit [25-26]. This limit, which was thought to be justified with the hope that the nonfactorizable part relative to the factorizable amplitude is of the order of $1/N_c$, was expected to perform even better for the *B*- meson decays, where the final state particles carry larger momenta than that of the charm meson decays. However, later the measurement of $\bar{B} \to D\pi$ meson decays did not favour this result empirically, as these decays require $a_1 \approx 1.03,$ and $a_2 \approx 0.23,$ *i.e.* a positive value of $a_2$, in sharp contrast to the expectations based on the large $N_c$ limit. Thus *B*- meson decays, revealing constructive interference between CF and CS diagrams for $B^- \to \pi^- D^0$, seem to favor $N_c = 3$ (real value). In our earlier work, using isospin analysis, we have searched for a systematics in the nonfactorizable contributions for various decays of $B^-$ and $\bar{B}^0$-mesons involving isospin l/2 and 3/2 final states. We observed that the nonfactorizable isospin 1/2 and 3/2 reduced amplitudes bear a universal ratio for $\bar{B} \to \pi D / \rho D / \pi D^*$ and $\bar{B} \to a_1 D / \pi D_1 / \pi D_1^{'} / \pi D_2 / \pi D_0$ decay modes [27-28]. In order to realize the full impact of the flavor symmetry, and to relate decays of strange bottom meson with those of non-strange bottom mesons and also to relate CKM-enhanced mode $(\Delta b = 1, \Delta C = 1, \Delta S = 0)$, with CKM-suppressed mode $(\Delta b = 1, \Delta C = 1, \Delta S = -1)$, we generalize our methodology to the *SU(3)* flavor symmetry for treatment of the nonfactorizable parts of corresponding weak decay amplitudes.

This paper is organized as follow. In section II, weak Hamiltonian is expressed as sum of two particle generating factorizable and nonfactorizable contributions to the weak hadronic decays of *B*-mesons. In section II A, we compute factorizable amplitudes and branching fractions; Tables 1 and 2 shows the whole set of data. In section II B, we illustrate brief methodology of our approach to calculate nonfatorizable amplitudes and presented these results in Tables 3 and 4 for CKM-enhanced and CKM-suppressed modes. We include our findings for total branching fractions in Tables 5 and 6 for comparison's sake. Section III discusses the results. Summary and conclusions are given in the last section.

## II. WEAK HAMILTONIAN

Structure for the general weak current $\otimes$ current Hamiltonian $H_w$ including short distance *QCD* effects for the CKM-enhanced mode $(\Delta b = 1, \Delta C = 1, \Delta S = 0)$,

$$H_w = \frac{G_F}{\sqrt{2}} V_{cb} V_{ud}^* \left[ c_1 \left( \bar{d}u \right) \left( \bar{c}b \right) + c_2 \left( \bar{c}u \right) \left( \bar{d}b \right) \right], \tag{1}$$

and for the CKM- suppressed mode $(\Delta b = 1, \Delta C = 1, \Delta S = -1)$,

$$H_w = \frac{G_F}{\sqrt{2}} V_{cb} V_{us}^* \left[ c_1 \left( \bar{s}u \right) \left( \bar{c}b \right) + c_2 \left( \bar{c}u \right) \left( \bar{s}b \right) \right], \tag{2}$$

where $\bar{q}_1 q_2 = \bar{q}_1\gamma_\mu(1-\gamma_5)q_2$ denotes color singlet $V–A$ Dirac current and the QCD coefficients [18] at bottom mass scale are

$$c_1 = 1.132, \qquad c_2 = -0.287. \tag{3}$$

The hadronic weak matrix elements of an operator, say $(\bar{d}u)(\bar{c}b)$, receive contributions from the operator itself and from its Fierz- transformed form:

$$(\bar{d}u)(\bar{c}b) = \frac{1}{N_c}(\bar{c}u)(\bar{d}b) + \frac{1}{2}\left(\bar{c}\lambda^c u\right)\left(\bar{d}\lambda^c b\right), \tag{4}$$

where $\bar{q}_1\lambda^c q_2 \equiv \bar{q}_1\gamma_\mu(1-\gamma_5)\lambda^c q_2$ represents the color octet current.

Finally the weak Hamiltonian (1) for CKM-enhanced mode acquires the following form:

$$H_w^{CF} = \frac{G_F}{\sqrt{2}}V_{cb}V_{ud}^*\left[a_1\left(\bar{d}u\right)_H\left(\bar{c}b\right)_H + c_2 H_w^8\right], \tag{5}$$

for color flavored (CF) diagrams, and

$$H_w^{CS} = \frac{G_F}{\sqrt{2}}V_{cb}V_{ud}^*\left[a_2\left(\bar{c}u\right)_H\left(\bar{d}b\right)_H + c_1\tilde{H}_w^8\right], \tag{6}$$

for color suppressed (CS) diagrams, where

$$a_{1,2} = c_{1,2} + \frac{c_{2,1}}{N_c},$$

$$H_w^8 = \frac{1}{2}\sum_{a=1}^{8}\left(\bar{c}\lambda^a u\right)\left(\bar{d}\lambda^a b\right), \qquad \tilde{H}_w^8 = \frac{1}{2}\sum_{a=1}^{8}\left(\bar{d}\lambda^a u\right)\left(\bar{c}\lambda^a b\right). \tag{7}$$

The subscript $H$ represents the change from quark currents to hadron field operators. Similar treatment can be done for the CKM- suppressed weak Hamiltonian (2).

The weak Hamiltonian thus generates factorizable (produced through color singlet weak current) and nonfactorizable parts (produced through color-octet weak currents) of the weak decay amplitudes, which can be expressed as

$$A^{tot}(\bar{B} \to P_1P_2) = A^f(\bar{B} \to P_1P_2) + A^{nf}(\bar{B} \to P_1P_2). \tag{8}$$

## A. EVALUATION OF FACTORIZABLE AMPLITUDES: $A^f(\bar{B} \to P_1P_2)$

Following the standard procedure to calculate the factorizable part in CKM-enhanced and CKM-suppressed modes, we obtain spectator quark decay amplitudes as given in col. 2 of Tables 1 and 2 respectively.

We calculate the factorization contributions numerically using $N_c = 3$ (real value), which fixes

$$a_1 = 1.02, \quad a_2 = +0.11. \tag{9}$$

Numerical inputs for the decay constants [29] are

$$f_D = (0.207 \pm 0.009)GeV, \quad f_\pi = (0.131 \pm 0.002)GeV,$$
$$f_K = (0.156 \pm 0.009)GeV. \tag{10}$$

Momentum dependence of the form- factors is taken as

$$F_0(q^2) = \frac{F_0(0)}{\left(1 - q^2/m_s^2\right)^n}, \tag{11}$$

where *n=1* for the monopole formula and pole mass is given by the scalar meson carrying the quantum number of the corresponding weak current [2-3], *i.e*,

$m_s = 6.80GeV$ for $(\bar{c}b)$ current,

$m_s = 5.89GeV$ for $(\bar{s}b)$ current,

$m_s = 5.78GeV$ for $(\bar{u}b)$ or $(\bar{d}b)$ current.

Form- factors at $q^2 = 0$ have the following values based on [4, 29-31]

$$F^{BK} = (0.34 \pm 0.00), \quad F^{BD} = (0.66 \pm 0.03), F^{B\pi} = (0.27 \pm 0.05),$$
$$F^{B\eta} = (0.22 \pm 0.05), F^{B\eta'} = (0.23 \pm 0.05),$$
$$F^{BsDs} = (0.67 \pm 0.01), F^{BsK} = (0.23 \pm 0.01),$$
$$F^{Bs\eta} = (0.24 \pm 0.01), F^{Bs\eta'} = (0.36 \pm 0.02). \quad (12)$$

There are many other calculations for these form-factors and decay constants, such as light-cone sum rules [30, 32], perturbative QCD approach [31, 33-34], FAT factorization assisted topological-amplitude approach [18], lattice QCD [35-36], LFCQM [29]. Incidentally, their estimates reasonably match with (10) and (12).

We use the $\eta$-$\eta$' mixing angle $\theta_p = \theta_{ideal} - \phi_P$, and $\phi_P = -15.4^\circ$ [1]. Using these numerical inputs, we calculate factorizable contributions as given in col. 3 of Table 1 for *CKM*-enhanced decays and in col. 3 of Table 2 for CKM- suppressed decays and the corresponding branching fractions are given in col. 4 of Tables 1 and 2 respectively.

The choice of $N_c = 3$ seems to work for $B^- \to D^0\pi^-$ and $B^- \to D^0 K^-$ which require constructive interference between *CF* and *CS* terms. However, other decays, especially those arising from the *CS* diagram, are experiencing large deviations from the experimental values. It is clear from the Tables 1 and 2 that the factorization alone is not sufficient to explain the measured branching fractions, which thus require nonfactorizable terms.

**Table 1. Spectator decay amplitudes (in col. 3) and branching fractions (in col. 4) for the CKM-enhanced decays (Δb = 1, ΔC = 1, ΔS = 0) are given. The last column gives the corresponding experimental branching fraction for comparison.**

| **Decay Process** | **Factorizable –Amplitudes ($A^f$)** $\left(\times G_F/\sqrt{2}\,GeV^3\right)$ | $A^f$ $\left(\times G_F/\sqrt{2}\,GeV^3\right)$ | $B^{fac}$ $\left(\times 10^{-4}\right)$ | $B^{Exp}$ $\left(\times 10^{-4}\right)$ |
|---|---|---|---|---|
| $B^- \to D^0 \pi^-$ | $a_1 f_\pi F_0^{BD}(m_\pi^2)(m_B^2 - m_D^2)$ $+ a_2 f_D F_0^{B\pi}(m_D^2)(m_B^2 - m_\pi^2)$ | 9.482 ± 0.421 | 50.51 ± 4.49 | 46.8±1.3 |
| $\bar{B}^0 \to D^0 \pi^0$ | $-a_2 f_D F_0^{B\pi}(m_D^2)(m_B^2 - m_\pi^2)/\sqrt{2}$ | -0.558 ± 0.103 | 0.17 ± 0.06 | 2.63±0.14 |
| $\bar{B}^0 \to D^+ \pi^-$ | $a_1 f_\pi F_0^{BD}(m_\pi^2)(m_B^2 - m_D^2)$ | 8.687 ± 0.395 | 40.06 ± 3.69 | 25.2±1.3 |
| $\bar{B}^0 \to D^0 \eta$ | $a_2 f_D F_0^{B\eta}(m_D^2)(m_B^2 - m_\eta^2)\sin\theta_p/\sqrt{2}$ | 0.586 ± 0.016 | 0.18 ± 0.01 | 2.36±0.32 |
| $\bar{B}^0 \to D^0 \eta'$ | $a_2 f_D F_0^{B\eta'}(m_D^2)(m_B^2 - m_{\eta'}^2)\cos\theta_p/\sqrt{2}$ | 0.322 ± 0.013 | 0.052±0.004 | 1.38±0.16 |
| $\bar{B}^0 \to D_s^+ K^-$ | 0 | 0 | 0 | 0.27±0.05 |
| $\bar{B}_s^0 \to D^0 K^0$ | $a_2 f_D F_0^{Bs\,K}(m_D^2)(m_{Bs}^2 - m_K^2)$ | 0.840 ± 0.030 | 0.36 ± 0.03 | --- |
| $\bar{B}_s^0 \to D_s^+ \pi^-$ | $a_1 f_\pi F_0^{Bs\,Ds}(m_\pi^2)(m_{Bs}^2 - m_{Ds}^2)$ | 8.759 ± 0.404 | 39.12 ± 3.60 | 30.0±2.3 |

**Table 2. Spectator decay amplitudes (in col. 3) and branching fractions (in col. 4) for the CKM-suppressed decays (Δb = 1, ΔC = 1, ΔS = -1) are given. The last col. gives the corresponding experimental branching fraction for comparison.**

| **Decay Process** | **Factorizable –Amplitudes ($A^f$)** $\left(\times G_F/\sqrt{2}\,GeV^3\right)$ | $A^f$ $\left(\times G_F/\sqrt{2}\,GeV^3\right)$ | $B^{fac}$ $\left(\times 10^{-4}\right)$ | $B^{Exp}$ $\left(\times 10^{-4}\right)$ |
|---|---|---|---|---|
| $B^- \to D^0 K^-$ | $a_2 f_D F_0^{BK}(m_D^2)(m_B^2 - m_K^2)$ $+ a_1 f_K F^{BD}(m_K^2)(m_B^2 - m_D^2)$ | 2.718±0.116 | 4.10 ±0.35 | 3.63±0.12 |
| $\bar{B}^0 \to D^0 \bar{K}^0$ | $a_2 f_D F_0^{BK}(m_D^2)(m_B^2 - m_K^2)$ | 0.277 ± 0.040 | 0.04 ± 0.01 | 0.52±0.07 |
| $\bar{B}^0 \to D^+ K^-$ | $a_1 f_K F_0^{BD}(m_K^2)(m_B^2 - m_D^2)$ | 2.439 ± 0.111 | 3.12 ± 0.28 | 1.86±0.2 |
| $\bar{B}_s^0 \to D_s^+ K^-$ | $a_1 f_K F_0^{BsDs}(m_K^2)(m_{Bs}^2 - m_{Ds}^2)$ | 2.459 ± 0.114 | 3.05 ± 0.28 | 2.27±0.19 |
| $\bar{B}_s^0 \to D^0 \pi^0$ | 0 | 0 | 0 | --- |
| $\bar{B}_s^0 \to D^+ \pi^-$ | 0 | 0 | 0 | --- |
| $\bar{B}_s^0 \to D^0 \eta$ | $-a_2 \cos\theta_p f_D F_0^{Bs\eta}(m_D^2)(m_{Bs}^2 - m_\eta^2)$ | -0.121 ± 0.009 | 0.0075 ± 0.0011 | --- |
| $\bar{B}_s^0 \to D^0 \eta'$ | $a_2 \sin\theta_p f_D F_0^{Bs\eta'}(m_D^2)(m_{Bs}^2 - m_{\eta'}^2)$ | 0.144 ± 0.010 | 0.010±0.001 | --- |

## B. EVALUATION OF NONFACTORIZABLE AMPLITUDE: $A^{nf}(\bar{B} \to P_1P_2)$

The nonfactorizable contributions arise through the Hamiltonian made up of color-octet currents, $\langle P_1P_2 | H_w^8 | B \rangle$ and $\langle P_1P_2 | \tilde{H}_w^8 | B \rangle$. Using the Wigner Eckart theorem, these are expressed in terms of the *SU(3)* Clebsch-Gordan *(C. G.)* coefficients and the reduced weak amplitudes

$$\langle P_1P_2 | H_w^8 | B \rangle = \sum_{\gamma} (C.G.) \langle P_1\, P_2 \| H_w^8 \| B \rangle_{\gamma}\,, \tag{13}$$

and

$$\langle P_1P_2 | \tilde{H}_w^8 | B \rangle = \sum_{\gamma} (C.G) \langle P_1\, P_2 \| \tilde{H}_w^8 \| B \rangle_{\gamma}\,, \tag{14}$$

where $\gamma$ represents different weak diagrams like *W*-external emission, *W*-internal emission, *W*-exchange, *W*-annihilation and *W*-loop.

However, we employ tensor approach to evaluate the *C.G.* Coefficients, where the matrix element $\langle P_1P_2 | H_w^8 | B \rangle$ is considered as a weak spurion + $B \to P + P$ scattering process, whose general structure can be written in the *SU(3)* symmetry framework as

$$\begin{aligned} &[a(B^m\, P^i_m\, P_n - B^m\, P_m\, P^i_n) + d(\,B^i\, P^m_n\, P_m - B^i\, P_n\, P^m_m\,) \\ &+ a'(B^m\, P^i_m\, P_n + B^m\, P_m\, P^i_n) + d'(B^i\, P^m_n\, P_m + B^i\, P_n\, P^m_m\,)]\, H^n_i\,, \end{aligned} \tag{15}$$

there exists a straight correspondence between the terms in (15) and various quark level processes. In the flavor *SU(4)* symmetry, Hamiltonian for weak processes belongs to the representation appearing in $4^* \otimes 1 \otimes 4^* \otimes 4 = 4^* \oplus 4^* \oplus 20' \oplus 36^*$, the *SU(3)* octet is contained in $20'$ and $36^*$, so the coefficients *a* and *d* belong to the octet of $20'$ and other coefficients *a'* and *d'* belongs to the different octet of $36^*$ [12-13]. In (15), $B^a$ denotes triplet of *B*-meson

$$B^a \equiv \left(B^-, \bar{B}^0, \bar{B}_s^0\right), \tag{16}$$

$P_m$ denotes triplet of charm mesons,

$$P_a = \left[D^0, D^+, D_s^+\right], \tag{17}$$

and $P_b^a$ denotes $3 \otimes 3$ matrix of bottomless and charmless octet mesons,

$$P_j^i = \begin{bmatrix} P_1^1 & \pi^+ & K^+ \\ \pi^- & P_2^2 & K^0 \\ K^- & \bar{K}^0 & P_3^3 \end{bmatrix}, \tag{18}$$

where

$$P_1^1 = \frac{\pi^0 + \eta \sin\theta_p + \eta' \cos\theta_p}{\sqrt{2}},$$
$$P_2^2 = \frac{-\pi^0 + \eta \sin\theta_p + \eta' \cos\theta_p}{\sqrt{2}},$$
$$P_3^3 = -\eta \cos\theta_p + \eta' \sin\theta_p.$$

$H_j^i$ represents transformation behavior of the weak Hamiltonian (1) and (2) through the following *SU(3)* decomposition:

$$1 \otimes 1 \otimes 3^* \otimes 3 = 8 \oplus 1. \tag{19}$$

Since all the quarks appearing in the Hamiltonian are different, singlet does not contribute. Choosing $H_2^1$ and $H_3^1$ components of the weak spurion in Hamiltonian (15), respectively for *CKM*- enhanced mode and *CKM*- suppressed modes, we obtain nonfactorizable contributions to various $\bar{B} \to PP$ decays, which are given in the col. 2 of Tables 3 and 4, where QCD coefficient $c_2$ and $c_1$ have been appropriately multiplied for CF and CS terms coming from (5) and (6) respectively. It is to be pointed out that *W*-loop and *W*-annihilation diagrams do not contribute to the decays considered in this work.

There exists a straight correspondence between the terms appearing in (15) and various quark level processes. The terms, involving the coefficients *(-d + d')*, represent *W*-exchange diagrams. Other terms, having coefficient *(a + a'*) represent *W*-external emission, *(-a + a')* represent *W*-internal emission, collectively called, spectator quark decay like diagrams, where the bottomless quark in the parent *B*-meson flows into one of the final state mesons [13, 37]. For the sake of clarity, we make the following substitution

$$E = a + a', I = -a + a', \text{ and } X = -d + d', \qquad (20)$$

corresponding to *W*-external emission, *W*-internal emission and *W*-exchange diagrams respectively. Consequently all the nonfactorizable terms involve only three reduced amplitudes to be fixed as shown in col. 3 of Tables 3 and 4.

As we know that the nonfactorizable contributions are not calculable exactly from theory at present, so these have to be estimated from the available data. In order to reduce the number of unknown parameters, we start by ignoring the *W*-exchange contribution, which has been found to be small. Specially, these may not play a significant role, when *W*- emission processes are contributing dominantly to a particular decay.

In order to evaluate these parameters, we choose those decay modes which are free from the FSI phases and *W*-exchange phenomenon. Using the experimental value of branching fraction for $B(B^- \to D^0\pi^-) = (4.68 \pm 0.13)\times 10^{-3}$, and subtracting the corresponding factorizable part from its experimental decay amplitude with a positive sign, we calculate

$$a' = -(0.05 \pm 0.02)\, GeV^3. \qquad (21)$$

Similarly, using the experimental value for $B(\bar{B}_s^0 \to D_s^+\pi^-) = (3.00 \pm 0.23)\times 10^{-3}$, we get

$$E = a + a' = (1.05 \pm 0.02)\, GeV^3, \qquad (22)$$

for positive sign of its amplitude, which leads to the following

$$I = -a + a' = (1.15 \pm 0.32)\, GeV^3, \tag{23}$$

Consequently, we calculate values of the nonfactorizable contributions for all the remaining decays, which are given in col.4 of Tables 3 and 4 for CKM- enhanced and CKM- suppressed decays, respectively.

**Table 3. Nonfactorizable contributions to $\bar{B} \to PP$ decays (Δb = 1, ΔC = 1, ΔS = 0) $\left(\times G_F/\sqrt{2}\, GeV^3\right)$**

| Decay Process | NonFactorizable-Amplitude $A^{nf}$ | NonFactorizable-Amplitude $A^{nf}$ | $A^{nf}$ ($\times 10^{-2}$) |
|---|---|---|---|
| $B^- \to D^0 \pi^-$ | $2a'(c_1 + c_2)$ | $(I+E)(c_1 + c_2)$ | -0.355±0.127 |
| $\bar{B}^0 \to D^0 \pi^0$ | $(a-a'-d+d')\, c_1/\sqrt{2}$ | $(-I+X)\,(1/\sqrt{2})\, c_1$ | 3.642±1.012 |
| $\bar{B}^0 \to D^+ \pi^-$ | $(a+a'-d+d')\, c_2$ | $(E+X) c_2$ | -1.088±0.332 |
| $\bar{B}^0 \to D^0 \eta$ | $(-a+a'-d+d') \sin\theta_p\; c_1/\sqrt{2}$ | $(I+X) \sin\theta_p\; c_1/\sqrt{2}$ | -2.818±0.783 |
| $\bar{B}^0 \to D^0 \eta'$ | $(-a+a'-d+d') \cos\theta_p\; c_1/\sqrt{2}$ | $(I+X) \cos\theta_p\; c_1/\sqrt{2}$ | -2.307±0.640 |
| $\bar{B}^0 \to D_s^+ K^-$ | $(-d+d')\, c_2$ | $X\, c_2$ | 0 |
| $\bar{B}_s^0 \to D^0 K^0$ | $(-a+a')\; c_1$ | $I\; c_1$ | -5.151±1.432 |
| $\bar{B}_s^0 \to D_s^+ \pi^-$ | $(a+a')\; c_2$ | $E\, c_2$ | -1.088±0.332 |

**Table 4. Nonfactorizable contributions to $\bar{B} \to PP$ decays (Δb = 1, ΔC = 1, ΔS = -1) $\left(\times G_F/\sqrt{2}\, GeV^3\right)$**

| Decay Process | NonFactorizable-Amplitude $A^{nf}$ | NonFactorizable-Amplitude $A^{nf}$ | $A^{nf}$ ($\times 10^{-2}$) |
|---|---|---|---|
| $B^- \to D^0 K^-$ | $2a'(c_1 + c_2)$ | $(I+E)\,(c_1 + c_2)$ | -0.081±0.029 |
| $\bar{B}^0 \to D^0\; \bar{K}^0$ | $(-a+a')\, c_1$ | $I\, c_1$ | -1.176±0.327 |
| $\bar{B}^0 \to D^+ K^-$ | $(a+a')\, c_2$ | $E\, c_2$ | -0.248±0.076 |
| $\bar{B}_s^0 \to D_s^+ K^-$ | $(a+a'-d+d')\, c_2$ | $(E+X)\, c_2$ | -0.248±0.076 |
| $\bar{B}_s^0 \to D^0 \pi^0$ | $(-d+d')\, c_2/\sqrt{2}$ | $X\, c_2/\sqrt{2}$ | --- |
| $\bar{B}_s^0 \to D^+ \pi^-$ | $(-d+d')\, c_2$ | $X\, c_2$ | --- |
| $\bar{B}_s^0 \to D^0 \eta$ | $((a- a') \cos\theta_p + (-d+d') \sin\theta_p /\sqrt{2})\, c_1$ | $(-I \cos\theta_p + X \sin\theta_p /\sqrt{2}) c_1$ | 0.745±0.207 |
| $\bar{B}_s^0 \to D^0 \eta'$ | $(2(-a+ a') \sin\theta_p + (-d+ d') \cos\theta_p /\sqrt{2})\, c_1$ | $(I \sin\theta_p + X \cos\theta_p /\sqrt{2})\, c_1$ | -0.910±0.253 |

It is to be noted that for the case of decays involving *CF* processes, nonfactorizable terms are small in comparison to the factorization terms. In contrast, for the decays appearing through the *CS* process, nonfactorizable terms are significantly large, bringing theory closer to the experiment.

We calculate $A\left(\bar{B} \to P_1 P_2\right)$ by adding $A^f$ (given in col. 3 of Tables 1 and 2) and $A^{nf}$ (given in col. 4 of Tables 3 and 4) for various decays. The branching fraction for *B*-meson decays into two pseudoscalar mesons is related to its decay amplitude as follows:

$$B\left(\bar{B} \to P_1 P_2\right) = \tau_B \left| \frac{G_F}{\sqrt{2}} V_{cb} V_{ud}^{*} \right|^2 \frac{p}{8\pi m_B^2} \left| A\left(\bar{B} \to P_1 P_2\right) \right|^2 , \qquad (24)$$

given in col. 3 of Tables 5 and 6, which are in better agreement than the case of factorization alone (given in col. 2 Tables 5 and 6) as compared to the experimental data available (given in col. 5 Tables 5 and 6). Our results also match with Chua [16] (given in col. 4 Tables 5 and 6).

**Table 5. Branching fractions of CKM-favored decays ($\Delta b = 1, \Delta C = 1, \Delta S = 0$)**

| Decay Process | Theoretical-Branching Only-factorization $B^f$ $(\times 10^{-4})$ | Theoretical-Branching $B^{Theo}$ $(\times 10^{-4})$ | Chua [16] $B$ $(\times 10^{-4})$ | Experimental - Branching (Exp-Br) [1] $B^{Exp}$ $(\times 10^{-4})$ |
|---|---|---|---|---|
| $B^- \to D^0 \pi^-$ | 50.51 ± 4.49 | 46.8 ± 1.3 | $48.4_{-0.8}^{+0.8}$ | 46.8 ± 1.3$^\dagger$ |
| $\bar{B}^0 \to D^0 \pi^0$ | 0.17 ± 0.06 | 5.05 ± 3.33 | $2.42_{-0.16}^{+0.19}$ | 2.63 ± 0.14 |
| $\bar{B}^0 \to D^+ \pi^-$ | 40.06 ± 3.69 | 30.7 ± 4.2 | $26.9_{-1.0}^{+1.0}$ | 26.8 ± 1.3 |
| $\bar{B}^0 \to D^0 \eta$ | 0.18 ± 0.01 | 2.61 ± 1.83 | $2.06_{-0.29}^{+0.30}$ | 2.36 ± 0.32 |
| $\bar{B}^0 \to D^0 \eta'$ | 0.052±0.004 | 1.99 ± 1.28 | $1.27_{-0.19}^{+0.21}$ | 1.38 ± 0.16 |
| $\bar{B}^0 \to D_s^+ K^-$ | 0 | 0 | 0.26 ± 0.03 | 0.22 ± 0.05 |
| $\bar{B}_s^0 \to D^0 K^0$ | 0.36 ± 0.03 | 9.5 ± 5.2 | --- | --- |
| $\bar{B}_s^0 \to D_s^+ \pi^-$ | 39.12 ± 3.60 | 30.0 ± 2.3 | --- | 30.0 ± 2.3$^\dagger$ |

$^\dagger$inputs

**Table 6. Branching fraction of CKM-suppressed decays ($\Delta b = 1, \Delta C = 1, \Delta S = -1$)**

| Decay Process | Theoretical-Branching Only-factorization $B^f$ $(\times 10^{-4})$ | Theoretical-Branching $B^{Theo}$ $(\times 10^{-4})$ | Chua [16] $B$ $(\times 10^{-4})$ | Experimental -Branching (Exp-Br) [1] $B^{Exp}$ $(\times 10^{-4})$ |
|---|---|---|---|---|
| $B^- \to D^0 K^-$ | 4.10 ±0.35 | 3.86 ± 0.35 | $4.01_{-0.09}^{+0.07}$ | 3.63 ± 0.12 |
| $\bar{B}^0 \to D^0 \bar{K}^0$ | 0.04 ± 0.01 | 0.42 ± 0.31 | $0.60_{-0.04}^{+0.03}$ | 0.52 ± 0.07 |
| $\bar{B}^0 \to D^+ K^-$ | 3.12 ± 0.28 | 2.52 ± 0.31 | 1.97 ± 0.07 | 1.86 ± 0.20 |
| $\bar{B}_s^0 \to D_s^+ K^-$ | 3.05 ± 0.28 | 2.46 ± 0.30 | --- | 2.27 ± 0.19 |
| $\bar{B}_s^0 \to D^0 \pi^0$ | 0 | --- | --- | --- |
| $\bar{B}_s^0 \to D^+ \pi^-$ | 0 | --- | --- | --- |
| $\bar{B}_s^0 \to D^0 \eta$ | 0.0075 ± 0.0011 | 0.20 ± 0.13 | --- | --- |
| $\bar{B}_s^0 \to D^0 \eta'$ | 0.010±0.001 | 0.29 ± 0.18 | --- | --- |

## III. RESULTS AND DISCUSSIONS

1) The branching fraction calculated by factorization alone overestimates for $B^- \to D^0\pi^-$ and $\bar{B}^0 \to D^+\pi^-$ decays. Though we have taken $B(B^- \to D^0\pi^-)$ as input, our results for $B(\bar{B}^0 \to D^+\pi^-)$ shows drastic improvement when nonfactorizable contributions are included.

2) For $\bar{B}^0 \to D^0\pi^0$ decay, we notice that the factorization alone underestimates its branching fraction, where as our value has the right order of magnitude. Here, we remind that these decays have shown the presence of FSI,

$$\begin{aligned} A(\bar{B}^0 \to D^+\pi^-) &= \frac{1}{\sqrt{3}}\left[A_{3/2}^{D\pi} e^{i\delta_{3/2}^{D\pi}} + \sqrt{2}A_{1/2}^{D\pi} e^{i\delta_{1/2}^{D\pi}}\right], \\ A(\bar{B}^0 \to D^0\pi^0) &= \frac{1}{\sqrt{3}}\left[\sqrt{2}A_{3/2}^{D\pi} e^{i\delta_{3/2}^{D\pi}} - A_{1/2}^{D\pi} e^{i\delta_{1/2}^{D\pi}}\right], \\ A(B^- \to D^0\pi^-) &= \sqrt{3}A_{3/2}^{D\pi} e^{i\delta_{3/2}^{D\pi}}. \end{aligned} \tag{25}$$

Following the FSI- phase independent analysis, sum of the branching fractions,

$$\begin{aligned} B\left(\bar{B}^0 \to D^+\pi^-\right) + B\left(\bar{B}^0 \to D^0\pi^0\right) &= \left(0.28 \pm 0.02\right)\% \quad \textit{Theo.} \\ &= \left(0.28 \pm 0.01\right)\% \quad \textit{Expt.} \end{aligned} \tag{26}$$

is in good agreement with the experiment.

3) We notice that factorization alone gives very small values of branching fraction for $\bar{B}^0 \to D^0\eta / D^0\eta'$ decays. However, our predictions are in nice agreement within experimental errors, when nonfactorizable terms are included.

4) We predict branching fraction for $\bar{B}_s^0 \to D^0 K^0$ decay to be $9.5\times10^{-4}$, which is 25 times larger than the prior value without nonfactorization. This presents litmus test for our scheme.

5) In the case of *CKM*- suppressed decay mode, $B^- \to D^0 K^-$ remains in agreement with the experimental value, and our prediction for $\bar{B}^0 \to D^0 \bar{K}^0$ gets significantly improved in comparison to the case of factorization alone. In fact, we notice that this decay occurs largely through the nonfactorizable terms.

6) It may also be noted that similar to $\bar{B} \to D\pi$ decays, $\bar{B} \to D\bar{K}$ decays are also subjected to *FSI* as their final states have two different Isospin states $I = 0$ and $I = 1$, which can evolve differently at the *B*-meson mass scale. Using the isospin analysis, we express their decay amplitudes as,

$$\begin{aligned} A(\bar{B}^0 \to D^+ K^-) &= \frac{1}{\sqrt{2}}\left[A_1^{DK} e^{i\delta_1} + A_0^{DK} e^{i\delta_0}\right], \\ A(\bar{B}^0 \to D^0 \bar{K}^0) &= \frac{1}{\sqrt{2}}\left[A_1^{DK} e^{i\delta_1} - A_0^{DK} e^{i\delta_0}\right], \qquad (27) \\ A(B^- \to D^0 K^-) &= \sqrt{2} A_1^{DK} e^{i\delta_1}, \end{aligned}$$

where $\delta_0$ and $\delta_1$ represent respective phases of $I = 0$ and 1 in the final state. Squaring and adding the first two equations, we calculate phase- independent sum of the branching fractions,

$$\begin{aligned} B\left(\bar{B}^0 \to D^+ K^-\right) + B\left(\bar{B}^0 \to D^0 \bar{K}^0\right) &= \left(2.94 \pm 0.42\right) \times 10^{-4} \quad \textit{Theo.} \\ &= \left(2.38 \pm 0.21\right) \times 10^{-4} \quad \textit{Expt.} \end{aligned} \qquad (28)$$

which shows better agreement within the errors.

7) Among the strange bottom mesons decays, our predicted branching fraction for $\bar{B}_s^0 \to D_s^+ K^-$ is in very nice agreement with the experimental value. Here, we notice that the nonfactorizable contributions show destructive interference with the facto- rizable amplitude.

8) We have also predicted branching fractions for, $\bar{B}_s^0 \to D^0\eta / D^0\eta'$ decays, which

are significantly larger than that of factorization alone. We also calculate branching fractions of, $\bar{B}^0 \to D^0\eta / D^0\eta'$ and $\bar{B}_s^0 \to D^0\eta / D^0\eta'$ decays for other values of the $\eta$-$\eta$' mixing angle, which are compared in Table 7. These can be tested in future experiments and present good tests of our analysis.

9) It is worth pointing out that the decays $\bar{B}^0 \to D_s^+ K^-, \bar{B}_s^0 \to D^0\pi^0$ and $\bar{B}_s^0 \to D^+\pi^-$ remain forbidden, as we have not included the *W*- exchange process. Though *W*-exchange factorizable contributions are expected to be highly suppressed due to the helicity and color arguments, its nonfactorizable counterpart may become noticeable due to the soft gluon exchange. Fixing the parameter *X* with $B\left(\bar{B}^0 \to D_s^+ K^-\right) = (0.27 \pm 0.05) \times 10^{-4}$, as input, we predict

$$\begin{aligned} B\left(\bar{B}_s^0 \to D^0\pi^0\right) &= (0.7 \pm 0.2) \times 10^{-6}, \\ B\left(\bar{B}_s^0 \to D^+\pi^-\right) &= (1.4 \pm 0.3) \times 10^{-6}, \end{aligned} \quad (29)$$

which may also be checked in future experiments.

**Table 7. Branching Fractions ( ×10$^{-4}$) of $\eta/\eta'$ emitting Decays including Nonfactorization Terms**

| Decay Process | $\phi$ = - 10.4° | $\phi$ = - 15.4° | $\phi$ = -24.4° | Expt-Branching [1] |
|---|---|---|---|---|
| $\bar{B}^0 \to D^0\eta$ | 2.23 ± 1.56 | 2.61 ± 1.83 | 3.24 ± 2.28 | 2.36 ± 0.32 |
| $\bar{B}^0 \to D^0\eta'$ | 2.43 ± 1.57 | 1.99 ± 1.28 | 1.27 ± 0.82 | 1.38 ± 0.16 |
| $\bar{B}_s^0 \to D^0\eta$ | 0.24 ± 0.16 | 0.20 ± 0.13 | 0.13 ± 0.08 | --- |
| $\bar{B}_s^0 \to D^0\eta'$ | 0.25 ± 0.16 | 0.29 ± 0.18 | 0.36 ± 0.24 | --- |

## IV. SUMMARY AND CONCLUSION

Bottom meson weak decays remain an exciting field for both theoretical and experimental investigations. Non-perturbated nonfactorizable contributions are difficult to calculate, from the theory of strong interactions, we employ *SU(3)* flavor symmetry to investigate the nonfactorizable contributions to CKM-enhanced and CKM-suppressed decays. In the section II A, we have calculated the branching fraction for factorization alone and found that it is not sufficient to explain the experimental branching fractions and then employed *SU(3)* flavor symmetry-framework for the nonfactorizable part. We have observed that there is drastic improvement in these results when nonfactorizable contributions are included. We also calculate branching fractions of, $\bar{B}^0 \to D^0\eta / D^0\eta'$ and $\bar{B}_s^0 \to D^0\eta / D^0\eta'$ decays for different values of the $\eta$-$\eta$' mixing angle, which are compared in Table 7. These can be tested in future experiments and present good tests of our analysis.